\documentclass[twocolumn]{article}
\usepackage[utf8]{inputenc}
\usepackage{microtype}
\usepackage[numbers,square]{natbib} 
\usepackage{color}
\usepackage{url}
\usepackage{amsmath, amssymb}
\usepackage{subcaption}
\usepackage{authblk}
\usepackage{graphicx}
\usepackage{fancyhdr}

\title{An Experimental Analysis of the Entanglement Problem in Neural-Network-based Music Transcription Systems}

\author[1]{Rainer Kelz \thanks{Address correspondence to: \texttt{rainer.kelz@jku.at}}}
\author[1]{Gerhard Widmer}

\affil[1]{Department of Computational Perception, Johannes Kepler University Linz, Austria}

\newcommand{\ConvNet}{\mbox{\textit{ConvNet}~}}
\newcommand{\SmallConvNet}{\mbox{\textit{SmallConvNet}~}}
\newcommand{\AUNet}{\mbox{\textit{AUNet}~}}

\newcommand{\ConvNetNPS}{\mbox{\textit{ConvNet}}}
\newcommand{\SmallConvNetNPS}{\mbox{\textit{SmallConvNet}}}
\newcommand{\AUNetNPS}{\mbox{\textit{AUNet}}}

\begin{document}
\maketitle
\thispagestyle{fancy}

\begin{abstract}
Several recent polyphonic music transcription systems have utilized deep neural networks to achieve state of the art results on various benchmark datasets, pushing the envelope on framewise and note-level performance measures. Unfortunately we can observe a sort of \textit{glass ceiling} effect. To investigate this effect, we provide a detailed analysis of the particular kinds of errors that state of the art deep neural transcription systems make, when trained and tested on a piano transcription task. We are ultimately forced to draw a rather disheartening conclusion: the networks seem to learn \textit{combinations} of notes, and have a hard time generalizing to unseen combinations of notes. Furthermore, we speculate on various means to alleviate this situation.

\end{abstract}

\section{Introduction}
The problem of polyphonic transcription can be formally described as the transformation of a time-ordered sequence of (audio) samples $\mathbf{X} = (\mathbf{x})_{t=0}^{T}, \mathbf{x}_t \in \mathcal{X}$ into a set of tuples $(t_{s}, t_{e}, F_0, A)$, describing start, end, fundamental frequency or pitch and optionally amplitude of the notes that were played. A slightly easier problem is framewise transcription, or tone-quantized multi-$\mathrm{F}_0$ estimation, where the output is a time-ordered sequence $\mathbf{Y} = (\mathbf{y})_{t=0}^{T}, \mathbf{y}_t \in \mathcal{Y}$, with $\mathbf{y}_t \in \{0, 1\}^{\mathrm{K}}$ being a vector of indicator variables and $\mathrm{K}$ denoting the tonal range. In other words, the $\mathbf{y}$ vectors specify the note pitches believed to be active in a given audio frame $\mathbf{x}$. Another simplifying assumption is usually the presence of only a single instrument, which more often than not turns out to be the piano, having a tonal range of $\mathrm{K} = 88$.

We will focus on framewise transcription systems only, as they turn out to be a crucial stage in the full transcription process, especially in so called \textit{hybrid systems} that post-process the framewise output with dynamic probabilistic models to extract the aforementioned tuples describing musical notes, such as \cite{Sigtia_Benetos_Boulanger_Weyde_Avila_Dixon_2015, Sigtia_Benetos_Dixon_2016}.

A diverse set of methods have been employed to tackle the framewise transcription problem, with non-negative matrix factorization being one of the more prominent methods.
Smaragdis and Brown with their seminal paper \cite{Smaragdis_Brown_2003} using non-negative matrix factorization (NMF) for polyphonic transcription already identified an undesirable property of the technique. NMF seeks to minimize the reconstruction error $\|\mathbf{X} - \mathbf{W}\mathbf{H}\|_{\mathrm{N}}$, where $\mathbf{X} \in \mathbb{R}_{+}^{D \times T}$ is the vector valued signal to reconstruct, $\mathbf{W} \in \mathbb{R}_{+}^{D \times d}$ is the dictionary, $\mathbf{H} \in \mathbb{R}_{+}^{d \times T}$ are the activations in time of the bases and $\mathrm{N}(\cdot)$ is a matrix norm. If no additional constraints are applied and no a priori knowledge is exploited, Smaragdis and Brown \cite{Smaragdis_Brown_2003} note that the method learns a dictionary of \textit{unique events}, rather than individual notes. Two remedies for this problem are also named: either choose sets of notes in such a way that from their intersection single notes can be identified, or present all individual notes in isolation, so a meaningful dictionary can be learned first. A similar effect is achievable if the dictionary matrix is harmonically constrained. The latter two methods seem to be popular choices in the literature \cite{Smaragdis_Brown_2003, Benetos_Ewert_Weyde_2014, Bertin_Badeau_Richard_2007, Bertin_Badeau_Vincent_2009, Dessein_Cont_Lemaitre_2010, Grindlay_Ellis_2009, OHanlon_Plumbley_2014, Vincent_Bertin_Badeau_2010, Weninger_Kirst_Schuller_Bungartz_2013, Khlif_Sethu_2015} to solve this problem for NMF.

We conduct a simple experiment to examine whether neural networks trained for a piano transcription task suffer from the same \textit{disentanglement} problems, followed by an analysis of two very different neural network architectures and the extent to which they exhibit this behavior.

\section{Methods}
Lacking proper theoretical analytic tools for the model class of neural networks, we resort to empirical tools, namely computational experiments. We train several deep neural networks in a supervised fashion for a framewise piano transcription task and analyze their error behavior. Adhering very closely to already established model architectures, as exemplified in \cite{Sigtia_Benetos_Dixon_2016, Kelz_Dorfer_Korzeniowski_Boeck_Arzt_Widmer_2016}, we deviate only in very few aspects, mostly concerning hyperparameter choices that affect training time but have little effect on performance. These parametrized functions we learn are of the following form: $f_{net}: \mathcal{X} \rightarrow \mathcal{Y}$, with $f_{net}$ in turn being composed of multiple simpler functions, commonly referred to as \textit{layers} in the neural network literature. An example of a network with an input, hidden and output layer would be $f_{net}(\mathbf{x}) = f_3(f_2(f_1(\mathbf{x}; \theta_3); \theta_2); \theta_3)$, where $f_i(\mathbf{z}_{i-1};\theta_i) = \sigma(\mathbf{W}_i \mathbf{z}_{i-1} + \mathbf{b}_i)$ with $\theta_i = \{\mathbf{W}_i, \mathbf{b}_i\}$ having matching dimensions to fit the output $\mathbf{z}_{i-1}$ of the previous layer. $\sigma(\cdot)$ is a nonlinear function applied elementwise. We note here that the functions $f_i$ may actually have more than one input $\textbf{z}$, and it may also be from layers other than the directly previous layer. We do not explicitly model convolution as it can be expressed as a matrix-matrix product, given $\textbf{W}$ and $\textbf{z}$ have the right shapes.

\begin{figure}[ht]
  \centering

  \begin{subfigure}[b]{0.2\textwidth}
  \includegraphics[scale=1.2]{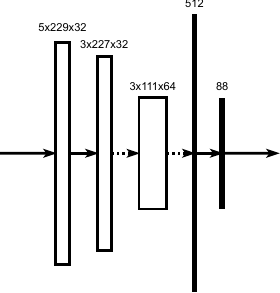}
  \caption{\ConvNet}
  \label{fig:drawing_convnet}
  \end{subfigure}
  ~
  \begin{subfigure}[b]{0.2\textwidth}
  \includegraphics[scale=1.2]{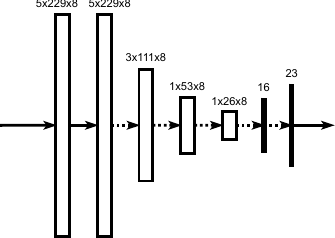}
  \caption{\SmallConvNet}
  \label{fig:drawing_small_convnet}
  \end{subfigure}

  \caption{The \ConvNet and \SmallConvNet architectures. White boxes denote convolutional layers, black thick lines stand for fully connected layers. Arrows show information flow, dashed lines indicate a max-pooling operation.}
\end{figure}

\begin{figure*}[ht]
  \centering
  \includegraphics[scale=1.0]{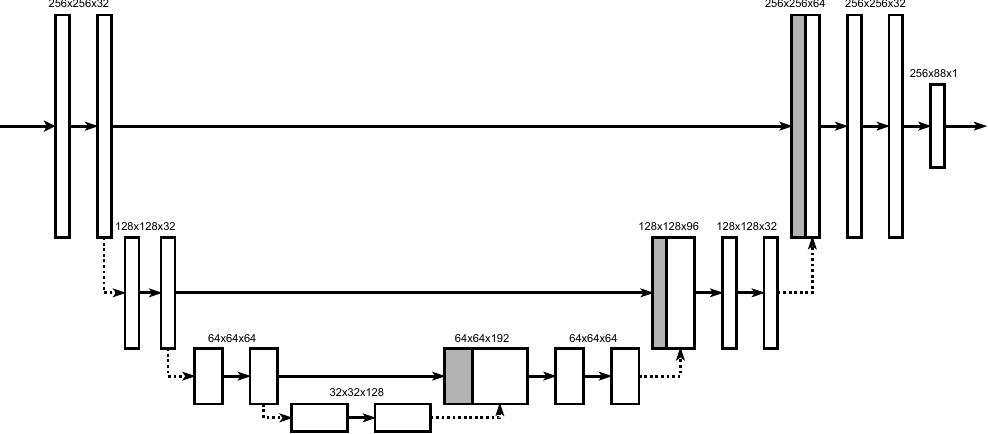}
  \caption{A schematic drawing of the \AUNet architecture. White boxes denote convolutional layers, their width corresponds to the number of convolutional kernels, their height corresponds to the size of the resulting feature maps. Arrows show information flow, dashed lines indicate either a max-pooling operation, if the line goes from a higher to a lower box, or an upscaling operation if the line goes from a lower to a higher box. Grey boxes next to white boxes denote a concatenation of feature maps. We made two adaptations to the original UNet architecture. The first is the use of upscaling operations instead of deconvolutions, and the second adaptation is the last layer having convolutions with a large kernel width in the frequency direction. \label{fig:drawing_unet}}
\end{figure*}

We choose neural network architectures already established to work well for framewise transcription. Our first choice is exactly the \ConvNet architecture as proposed in \cite{Kelz_Dorfer_Korzeniowski_Boeck_Arzt_Widmer_2016}, which achieves state of the art results for framewise transcription on a popular benchmark dataset. We also designed a much smaller version of this network, which will be referred to as \SmallConvNetNPS. Additionally, we borrow an architecture originally employed for medical image segmentation, called the UNet \cite{Ronneberger_Fischer_Brox_2015} and make two small modifications to adapt it for our purposes. We call the adapted architecture \AUNetNPS. It is able to directly integrate information at different scales, which is beneficial for smoothing in the temporal direction, and identifying groups of partials and their distance in the frequency dimension. The precise definitions for all networks, as well as schematic drawings of the architectures for the \ConvNetNPS, the \SmallConvNet and the \AUNet can be found in the appendix. The definitions are listed in tables \ref{table:convnet}, \ref{table:small_convnet} and \ref{table:aunet} whereas the schemata are depicted in figures \ref{fig:drawing_convnet}, \ref{fig:drawing_small_convnet} and \ref{fig:drawing_unet} respectively.

\section{Datasets}
We use a synthetic dataset to conduct small scale experiments with the \SmallConvNetNPS. A subset of the MAPS dataset \cite{Emiya_Badeau_David_2010} is used to train and test the \ConvNet and the \AUNet models. The MAPS dataset consists of several classical piano pieces, along with isolated notes and common chords, rendered with 7 different software synthesizers (samplers) in addition to 2 Disklavier piano recordings, one with the microphone close to the piano, and one with the microphone farther away and thus containing room acoustics. We now describe each subset in turn:

\subsection{FLUID}
For focused computational experiments we synthesize two-note combinations and isolated notes. We only use notes within an $11$ semitone range around a reference pitch (C4/MIDI60), creating $\binom{23}{2} = 253$ two-note intervals. The onset and offset of the two notes are exactly synchronous. We use the free software sampler Fluidsynth\footnote{\url{www.fluidsynth.org}} together with the freely available Fluid-R3-GM\footnote{\url{http://www.musescore.org/download/fluid-soundfont.tar.gz}} soundfont to render a dataset FLUID-COMBI where the train and validation sets both consist of the aforementioned intervals, whereas the test set contains individual notes only. For FLUID-ISOL, the individual notes are in the train and validation sets, whereas the test set contains the intervals. So for both datasets the intersection of unique events in train and test sets is the empty set. The error behavior of the \SmallConvNet on this dataset is discussed in section \ref{sec:results_fluid}.

\subsection{MAPS-MUS}
This subset consists only of the rendered classical piano pieces in the MAPS dataset. We adopt the more realistic train-test scenario described in \cite{Sigtia_Benetos_Dixon_2016}, which is referred to as \textit{Configuration-II}. It is more realistic because it trains only on synthetic renderings, and tests on the real piano recordings. We select the training set as all pieces from 6 synthesizers, the validation set is comprised of all renderings from a randomly selected 7th, and the test set is made up of all Disklavier recordings. We will refer to this dataset as MAPS-MUS from now on. The respective error behaviors of the two larger models, the \ConvNet and the \AUNet on this dataset are discussed in section \ref{sec:results_maps_mus}. We did not use the test set for conducting any error analysis, other than measuring final performance after model selection, to make sure that both models actually achieve state of the art results. The rationale behind this is explained in detail in section \ref{sec:results_maps_mus}.

\section{Results}
\subsection{\SmallConvNet and FLUID}
\label{sec:results_fluid}
We start with a controlled empirical analysis of the \textit{disentanglement} problem using our synthetic datasets. We train the \SmallConvNet for framewise transcription on logarithmic filtered, log-magnitude spectrograms with $229$ bins, as proposed in \cite{Kelz_Dorfer_Korzeniowski_Boeck_Arzt_Widmer_2016}. The output size of the network is limited to $23$ notes, and it has only $5327$ parameters, to make it approximately comparable to NMF with a dictionary matrix $\textbf{W} \in \mathbb{R}^{229 \times 23}$ having $5267$ parameters. We note that overfitting, fitting noise in the data, is not the real problem here, as the acoustic properties of the sound sources are the same for train and test set.

\begin{figure}[ht]
  \centering
  \includegraphics[scale=0.5]{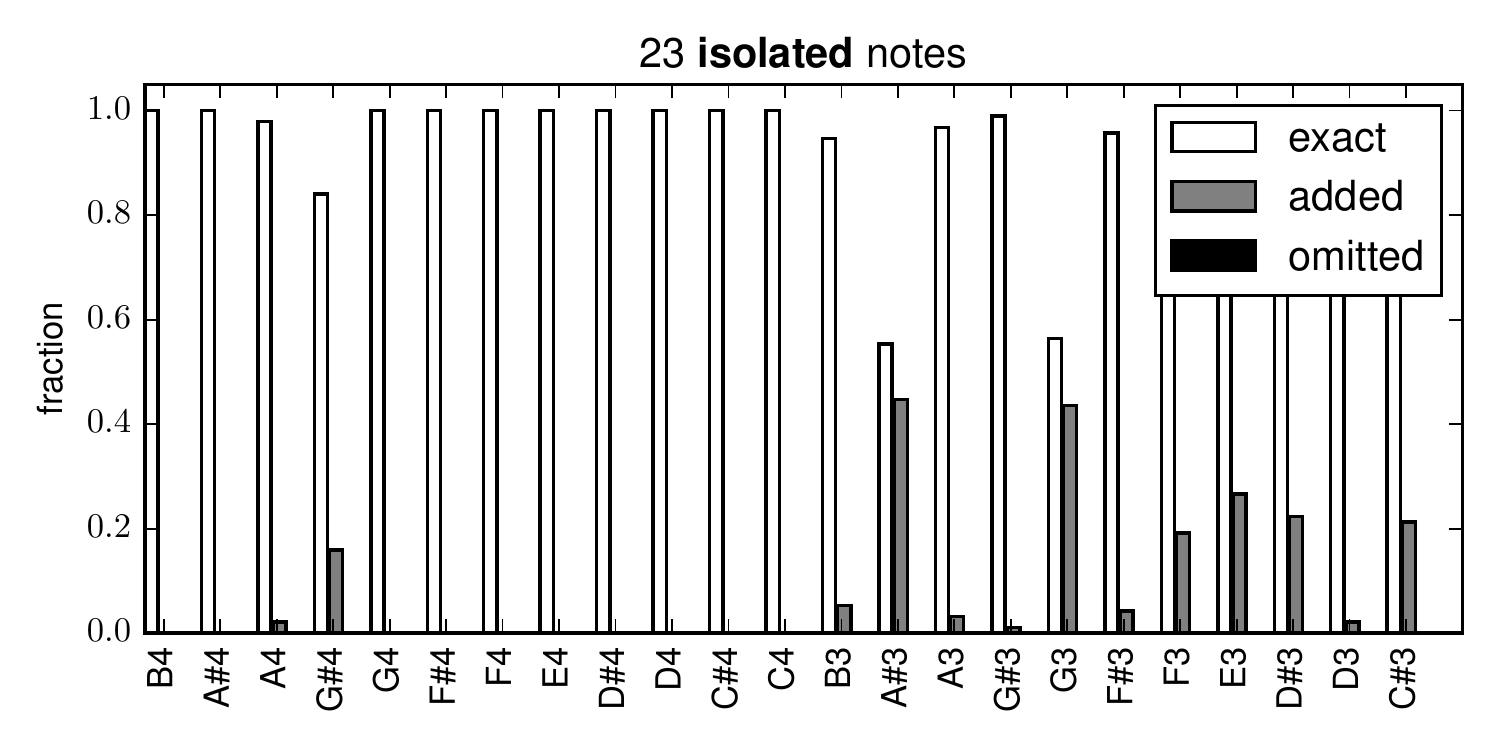}
  \caption{For isolated notes present \textbf{only} in the test set, this is the proportion of exactly transcribed frames, along with the proportions of frames that had notes added or omitted, respectively. Transcriptions stem from the \SmallConvNet trained on FLUID-COMBI. \label{fig:smallconv_fluid-combi}}
\end{figure}

\begin{figure}[ht]
  \centering
  \includegraphics[scale=0.5]{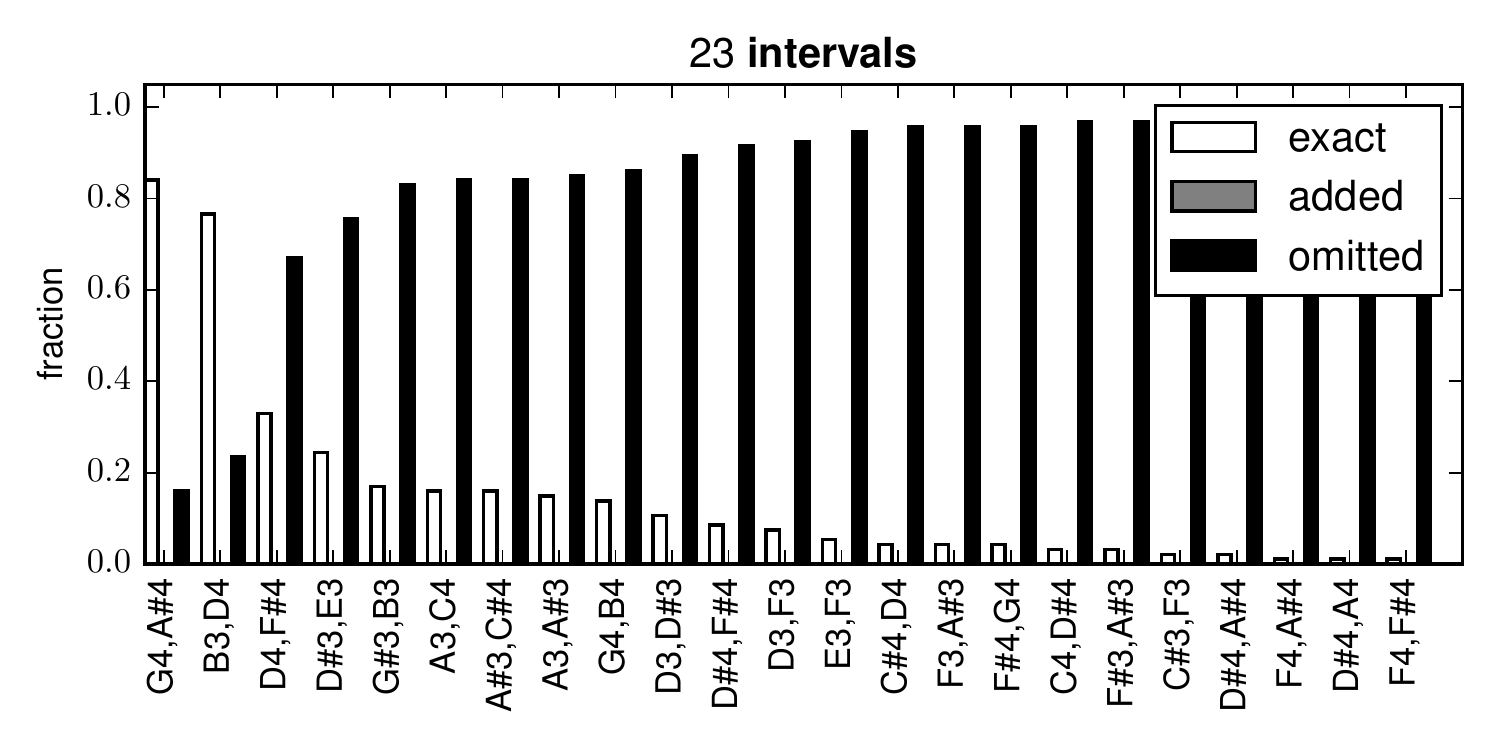}
  \caption{For a selection of the $23$ best transcribed intervals present \textbf{only} in the test set, this is the proportion of exactly transcribed frames, along with the proportions of frames that had notes added or omitted, respectively. Transcriptions stem from the  \SmallConvNet trained on FLUID-ISOL. \label{fig:smallconv_fluid-isol}}
\end{figure}

The general idea of this experiment is discovering to which extent the network is capable of detecting isolated notes, if all it has ever seen were combinations, and vice versa. We can find a partial answer to this question in figures \ref{fig:smallconv_fluid-combi} and \ref{fig:smallconv_fluid-isol}. The figures all show the proportion of frames where all notes have been exactly identified, and contrast them with the proportion of frames in which notes have been added or omitted. This means that the three quantities do not necessarily sum to one, because notes could have been added \textit{and} some others omitted in a frame.

In figure \ref{fig:smallconv_fluid-combi} we can observe that after seeing only \mbox{two-note} intervals, the network is able to generalize to isolated notes to some extent. While a surprising number of individual notes are transcribed perfectly, some notes are still not recognized properly. For these notes their companion notes from the train set are predicted as simultaneously sounding, indicating a failure to disentangle note combinations during training.

In figure \ref{fig:smallconv_fluid-isol} we see that the network utterly fails to generalize from isolated notes to note combinations, with only two exceptions. We plotted only the $23$ best transcribed note combinations, as for the remaining $230$ intervals the proportion of omission errors is very close to or even exactly $1.0$. The network does manage to transcribe two of the intervals with acceptable accuracy, however an explanation of why exactly these two intervals could be recognized eludes us at the moment.

We might draw a preliminary conclusion from these results: the strategy most successful for alleviating the \textit{disentanglement} problem for NMF, namely learning the dictionary $\textbf{W}$ from isolated notes, does not work for neural transcription systems. The NMF of spectrograms is a linear system, and therefore has the superposition property. Its response to multiple inputs is the sum of the responses for individual inputs. This is not necessarily true for neural networks, as they \textit{may} learn to approximate a linear function, but do not \textit{have} to.


The other strategy mentioned in \cite{Smaragdis_Brown_2003}, namely showing combinations of notes to the networks, seems to work fairly well for the majority of isolated notes, as can be observed in figure \ref{fig:smallconv_fluid-combi}. Unfortunately, the number of combinations for the tonal range of the piano grows large very quickly. Even when assuming a maximum polyphony of only $6$, we would already need to show $\sum_{i=2}^{6} \binom{88}{i} = 41.621.206$ combinations to the network.

\subsection{\ConvNetNPS, \AUNet and MAPS-MUS}
\label{sec:results_maps_mus}
We now turn our attention to a more musically relevant dataset. We train several instances of both a \ConvNet and an \AUNetNPS, closely adhering to the training procedure described in \cite{Kelz_Dorfer_Korzeniowski_Boeck_Arzt_Widmer_2016}, and select the model for analysis that achieves highest framewise f-measure on the validation set. Our analysis of error behavior is restricted to the validation set as well, simply because we want to avoid learning too much about the composition of the MAPS test set. The scenario is the same, as the validation set consists of pieces rendered by an unseen synthesizer. We feel that this also lends some additional strength to our argument, as we conduct our analysis on the best performing model for this set.

Two different scenarios are considered.
The first scenario looks at the transcription results for notes and note combinations that are present in both the train and validation set, referred to as ``shared'' combinations. A low proportion of additions will tell us that there were a sufficient number of examples for this particular combination, so it could not be overshadowed by combinations containing additional notes. A high proportion of omissions will indicate issues with generalization to different acoustic properties. If both proportions are high, this indicates that one or more notes in the combination have been mistaken for others.

The second scenario examines the transcription results for notes and note combinations that are present only in the validation set, referred to as ``unshared''. If the proportion of exactly transcribed frames is high, the network must have learned to disentangle individual notes from different combinations shown to it, and be able to recognize these disentangled parts in new, unseen combinations. A high proportion of additions will mainly tell us that the network has failed to disentangle parts, but still tries to combine the ones it knows about. A high proportion of omissions points to either a failure to simultaneously disentangle and recombine, a failure to generalize to different acoustic properties, or more probably both.

\begin{figure}[ht]
  \centering
  \includegraphics[scale=0.5]{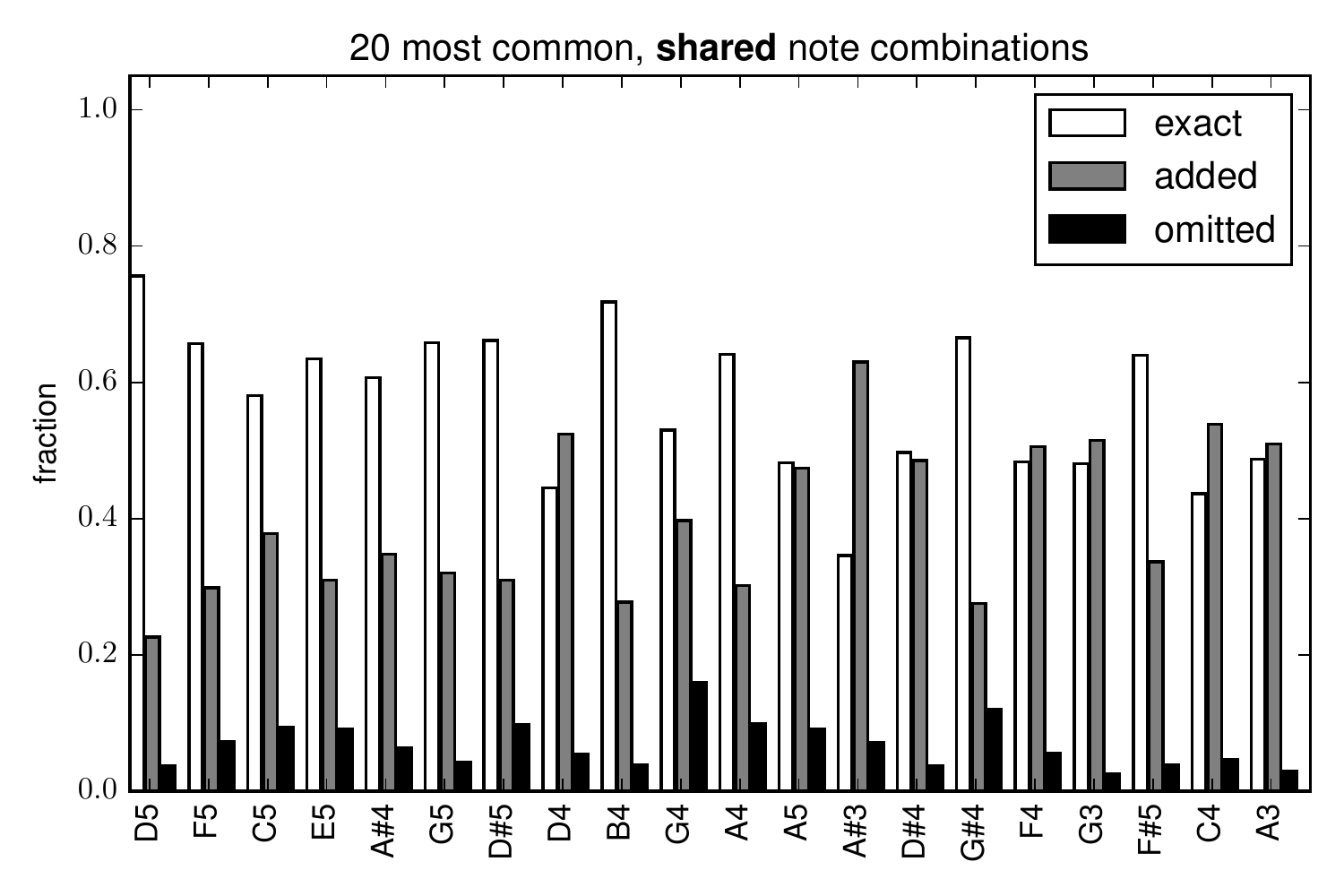}
  \caption{For the most common note combinations present \textbf{both} in the train set and validation set, this is the proportion of exactly transcribed frames, along with the proportion of frames that had notes added or omitted, respectively. Transcriptions stem from the \ConvNet trained on MAPS-MUS \label{fig:convnet-maps-mus-shared}}
\end{figure}

In figure \ref{fig:convnet-maps-mus-shared} we can see two things: the most common note combinations present in both train and validation set are actually isolated notes, and the relative frequency of exactly transcribed notes is comparatively high. Unfortunately, we can also see that the proportion of frames in which additional notes were erroneously transcribed is much higher than we would prefer, pointing to both a lack of examples for these individual notes at train time and the failure to generalize from combinations. They all are confused with combinations every so often. The low proportion of omission errors for isolated notes indicate only mild difficulties to generalize to different acoustical properties.

\begin{figure}[ht]
  \centering
  \includegraphics[scale=0.5]{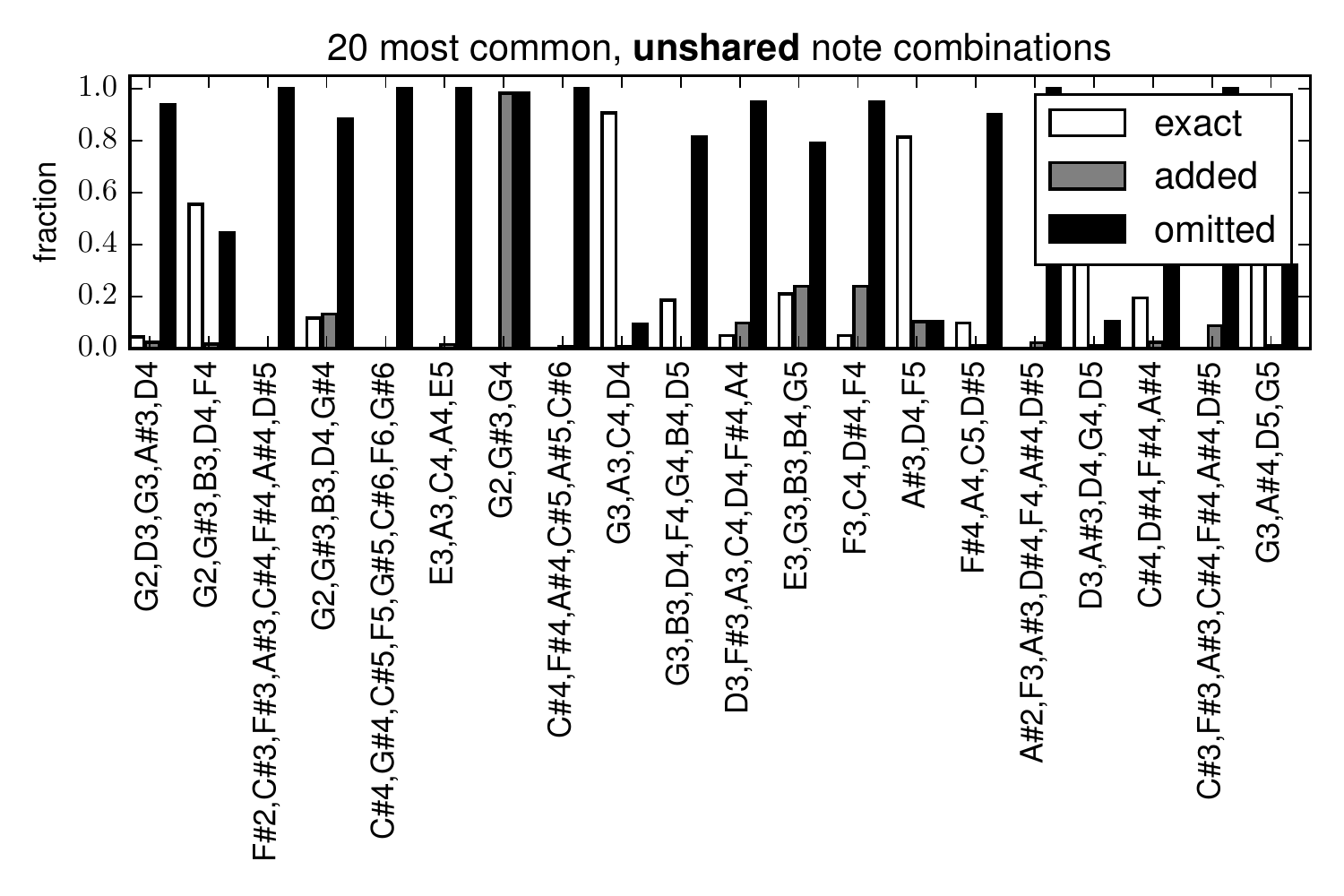}
  \caption{The most common note combinations present \textbf{only} in the validation set, and the proportion of exactly transcribed frames, along with the proportion of frames that had notes added or omitted, respectively. Transcriptions stem from the \ConvNet trained on MAPS-MUS. \label{fig:convnet-maps-mus-unshared}}
\end{figure}

Looking at figure \ref{fig:convnet-maps-mus-unshared}, we can see the error behavior of the network for the most common note combinations that are only present in the validation set. We notice a large amount of omission errors - which also indicates a failure to generalize to unseen note combinations. A few combinations, such as (G3, A3, C4, D4), stand out though as being transcribed with great accuracy. We could find no satisfactory explanation for this so far, other than the suspicion it has to do with their low polyphony.

\begin{figure}[ht]
  \centering
  \includegraphics[scale=0.5]{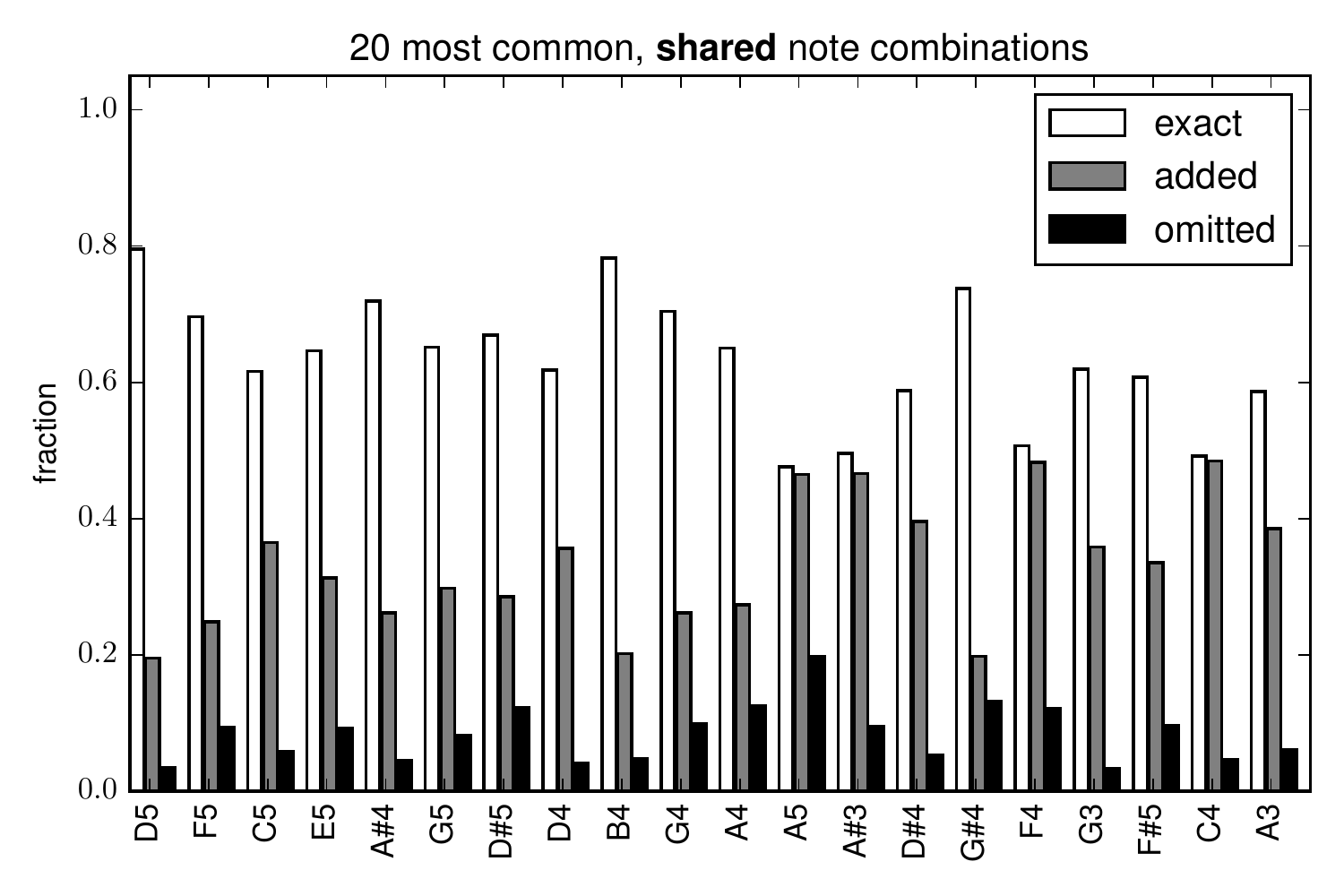}
  \caption{The most common note combinations present \textbf{both} in the train set and validation set, and proportion of exactly transcribed frames, along with the proportion of frames that had notes added or omitted, respectively. Transcriptions stem from the \AUNet trained on MAPS-MUS. \label{fig:unet-maps-mus-shared}}
\end{figure}

\begin{figure}[ht!]
  \centering
  \includegraphics[scale=0.5]{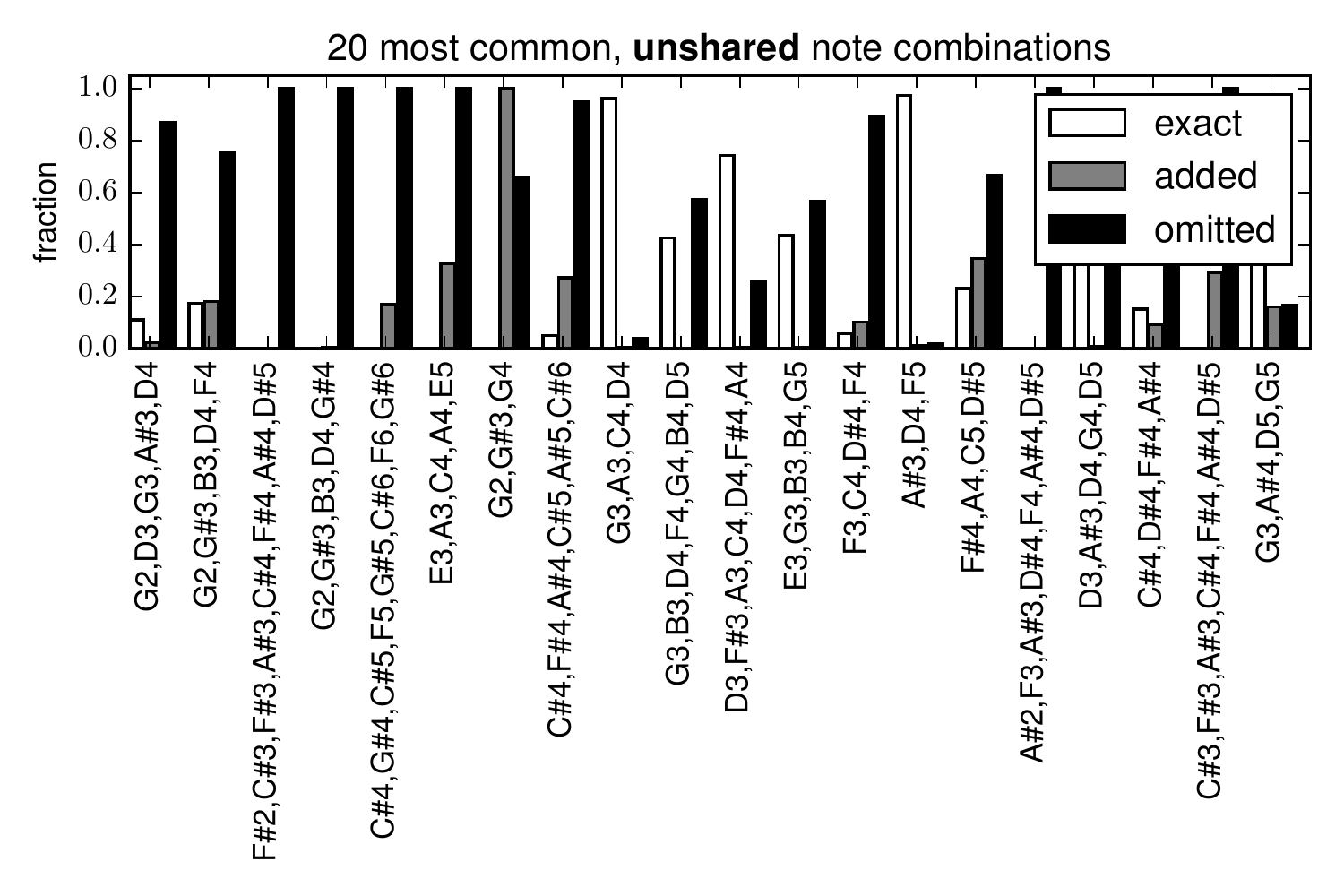}
  \caption{The most common note combinations present \textbf{only} in the validation set, and the proportion of exactly transcribed frames, along with the proportion of frames that had notes added or omitted, respectively. Transcriptions stem from the \AUNet trained on \mbox{MAPS-MUS}. \label{fig:unet-maps-mus-unshared}}
\end{figure}

If we compare the results of the \ConvNet (\mbox{figure \ref{fig:convnet-maps-mus-shared}}) and the \AUNet (\mbox{figure \ref{fig:unet-maps-mus-shared}}) for the most common note combinations which are shared by the train and validation set, we can observe that the \AUNet achieves marginally better exact transcription results across the board. In some cases, the proportion of added notes is reduced, however this happens at the expense of a slightly increased amount of omitted note combinations. Likewise, the results for the \ConvNet (\mbox{figure \ref{fig:convnet-maps-mus-unshared}}) and \AUNet transcriptions (\mbox{figure \ref{fig:unet-maps-mus-unshared}}) for the ``unshared'' case appear to be very similar, indicating a comparable error behavior across very different architectures.

Concluding this section we would like to emphasize that both architectures achieve the same (or even slightly exceed) framewise transcription results on the MAPS dataset as reported in \cite{Kelz_Dorfer_Korzeniowski_Boeck_Arzt_Widmer_2016}, which currently defines the state of the art. In other words, it is unlikely that the problematic results reported above are due to the fact that we made poor hyperparameter choices.

\section{Summary} 
We have experimentally shown that certain neural network architectures have difficulties \textit{disentangling} inputs which are superpositions or mixtures of individual parts, as discussed in section \ref{sec:results_maps_mus}. They learn to do so only if they are shown a large number of combinations whose constituent parts overlap, and they utterly fail to generalize to combinations when trained on individual parts of the mixture alone, as we determined in a small experiment described in section \ref{sec:results_fluid}. Any approach that tries to learn from a fixed set of combinations, for example defined by a set of music pieces, without incorporating additional constraints or prior knowledge, as is done in \cite{Sigtia_Benetos_Boulanger_Weyde_Avila_Dixon_2015, Sigtia_Benetos_Dixon_2016, Kelz_Dorfer_Korzeniowski_Boeck_Arzt_Widmer_2016}, will suffer from this problem.

The brute force approach to solve the \textit{disentanglement} problem would be showing all possible combinations to the network. Unfortunately this solution is intractable, due to the large tonal range and maximum polyphony of certain instruments. Arguably this approach would also not necessarily force the networks to learn how to \textit{disentangle}, as they could, in principle, simply memorize all combinations.
Learning a different note detector for each note, as done in \cite{Marolt_2004, Nam_Ngiam_Lee_Slaney_2011} suffers from the same problems, if the combinations shown to each detector are not diverse enough. Depending on the expressiveness of the model class, ``diverse enough'' could easily mean ``all combinations''.

A partial solution to this problem might involve a modification of the loss function for the network. An additional objective must specify the need to \textit{disentangle} individual notes explicitly. The network needs to learn to decompose a (nonlinear) mixture of signals into its constituent parts - a task commonly known as ``source separation''. Finding a formulation of a joint objective combining multi-label losses with a separation encouraging penalty that solves this \textit{disentanglement problem} is the topic of ongoing research.

\section*{Acknowledgements}
This work is supported by the European Research Council (ERC Grant Agreement 670035, project \mbox{CON ESPRESSIONE}). The Tesla K40 used for this research was donated by the NVIDIA Corporation.

\bibliographystyle{plain}
\bibliography{master}

\begin{thebibliography}{10}

\bibitem{Benetos_Ewert_Weyde_2014}
Emmanouil Benetos, Sebastian Ewert, and Tillman Weyde.
\newblock Automatic transcription of pitched and unpitched sounds from
  polyphonic music.
\newblock In {\em {IEEE} International Conference on Acoustics, Speech and
  Signal Processing, {ICASSP} 2014, Florence, Italy, May 4-9, 2014}, pages
  3107--3111, 2014.

\bibitem{Bertin_Badeau_Richard_2007}
Nancy Bertin, Roland Badeau, and Ga{\"{e}}l Richard.
\newblock Blind signal decompositions for automatic transcription of polyphonic
  music: {NMF} and {K-SVD} on the benchmark.
\newblock In {\em Proceedings of the {IEEE} International Conference on
  Acoustics, Speech, and Signal Processing, {ICASSP} 2007, Honolulu, Hawaii,
  USA, April 15-20, 2007}, pages 65--68, 2007.

\bibitem{Bertin_Badeau_Vincent_2009}
Nancy Bertin, Roland Badeau, and Emmanuel Vincent.
\newblock Fast bayesian nmf algorithms enforcing harmonicity and temporal
  continuity in polyphonic music transcription.
\newblock In {\em {IEEE} Workshop on Applications of Signal Processing to Audio
  and Acoustics, {WASPAA} '09, New Paltz, NY, USA, October 18-21, 2009}, pages
  29--32, 2009.

\bibitem{Dessein_Cont_Lemaitre_2010}
Arnaud Dessein, Arshia Cont, and Guillaume Lemaitre.
\newblock Real-time polyphonic music transcription with non-negative matrix
  factorization and beta-divergence.
\newblock In {\em Proceedings of the 11th International Society for Music
  Information Retrieval Conference, {ISMIR} 2010, Utrecht, Netherlands, August
  9-13, 2010}, pages 489--494, 2010.

\bibitem{Emiya_Badeau_David_2010}
Valentin Emiya, Roland Badeau, and Bertrand David.
\newblock Multipitch estimation of piano sounds using a new probabilistic
  spectral smoothness principle.
\newblock {\em {IEEE} Trans. Audio, Speech {\&} Language Processing},
  18(6):1643--1654, 2010.

\bibitem{Grindlay_Ellis_2009}
Graham Grindlay and Daniel P.~W. Ellis.
\newblock Multi-voice polyphonic music transcription using eigeninstruments.
\newblock In {\em {IEEE} Workshop on Applications of Signal Processing to Audio
  and Acoustics, {WASPAA} '09, New Paltz, NY, USA, October 18-21, 2009}, pages
  53--56, 2009.

\bibitem{Kelz_Dorfer_Korzeniowski_Boeck_Arzt_Widmer_2016}
Rainer Kelz, Matthias Dorfer, Filip Korzeniowski, Sebastian B{\"{o}}ck, Andreas
  Arzt, and Gerhard Widmer.
\newblock On the potential of simple framewise approaches to piano
  transcription.
\newblock In {\em Proceedings of the 17th International Society for Music
  Information Retrieval Conference, {ISMIR} 2016, New York City, United States,
  August 7-11, 2016}, pages 475--481, 2016.

\bibitem{Khlif_Sethu_2015}
Anis Khlif and Vidhyasaharan Sethu.
\newblock An iterative multi range non-negative matrix factorization algorithm
  for polyphonic music transcription.
\newblock In {\em Proceedings of the 16th International Society for Music
  Information Retrieval Conference, {ISMIR} 2015, M{\'{a}}laga, Spain, October
  26-30, 2015}, pages 330--335, 2015.

\bibitem{Marolt_2004}
Matija Marolt.
\newblock A connectionist approach to automatic transcription of polyphonic
  piano music.
\newblock {\em {IEEE} Trans. Multimedia}, 6(3):439--449, 2004.

\bibitem{Nam_Ngiam_Lee_Slaney_2011}
Juhan Nam, Jiquan Ngiam, Honglak Lee, and Malcolm Slaney.
\newblock A classification-based polyphonic piano transcription approach using
  learned feature representations.
\newblock In {\em Proceedings of the 12th International Society for Music
  Information Retrieval Conference, {ISMIR} 2011, Miami, Florida, USA, October
  24-28, 2011}, pages 175--180, 2011.

\bibitem{OHanlon_Plumbley_2014}
Ken O'Hanlon and Mark~D. Plumbley.
\newblock Polyphonic piano transcription using non-negative matrix
  factorisation with group sparsity.
\newblock In {\em {IEEE} International Conference on Acoustics, Speech and
  Signal Processing, {ICASSP} 2014, Florence, Italy, May 4-9, 2014}, pages
  3112--3116, 2014.

\bibitem{Ronneberger_Fischer_Brox_2015}
Olaf Ronneberger, Philipp Fischer, and Thomas Brox.
\newblock U-net: Convolutional networks for biomedical image segmentation.
\newblock In {\em Medical Image Computing and Computer-Assisted Intervention -
  {MICCAI} 2015 - 18th International Conference Munich, Germany, October 5 - 9,
  2015, Proceedings, Part {III}}, pages 234--241, 2015.

\bibitem{Sigtia_Benetos_Boulanger_Weyde_Avila_Dixon_2015}
Siddharth Sigtia, Emmanouil Benetos, Nicolas Boulanger{-}Lewandowski, Tillman
  Weyde, Artur~S. d'Avila Garcez, and Simon Dixon.
\newblock A hybrid recurrent neural network for music transcription.
\newblock In {\em 2015 {IEEE} International Conference on Acoustics, Speech and
  Signal Processing, {ICASSP} 2015, South Brisbane, Queensland, Australia,
  April 19-24, 2015}, pages 2061--2065, 2015.

\bibitem{Sigtia_Benetos_Dixon_2016}
Siddharth Sigtia, Emmanouil Benetos, and Simon Dixon.
\newblock An end-to-end neural network for polyphonic piano music
  transcription.
\newblock {\em {IEEE/ACM} Trans. Audio, Speech {\&} Language Processing},
  24(5):927--939, 2016.

\bibitem{Smaragdis_Brown_2003}
Paris Smaragdis and Judith~C. Brown.
\newblock Non-negative matrix factorization for polyphonic music transcription.
\newblock In {\em 2003 IEEE Workshop on Applications of Signal Processing to
  Audio and Acoustics (IEEE Cat. No. 03TH8684)}, page 177–180. IEEE, 2003.

\bibitem{Vincent_Bertin_Badeau_2010}
Emmanuel Vincent, Nancy Bertin, and Roland Badeau.
\newblock Adaptive harmonic spectral decomposition for multiple pitch
  estimation.
\newblock {\em {IEEE} Trans. Audio, Speech {\&} Language Processing},
  18(3):528--537, 2010.

\bibitem{Weninger_Kirst_Schuller_Bungartz_2013}
Felix Weninger, Christian Kirst, Bj{\"{o}}rn~W. Schuller, and Hans{-}Joachim
  Bungartz.
\newblock A discriminative approach to polyphonic piano note transcription
  using supervised non-negative matrix factorization.
\newblock In {\em {IEEE} International Conference on Acoustics, Speech and
  Signal Processing, {ICASSP} 2013, Vancouver, BC, Canada, May 26-31, 2013},
  pages 6--10, 2013.

\end{thebibliography}

\newpage
\section*{Appendix}

\begin{table}[htp]
\centering
\scalebox{0.6}{%
\begin{tabular}{llr}
	Layer                & Output          & No. of          \\ 
	Type                 & Dimensions      & Params          \\ 
\hline
	Input                & 1x5x229         &                 \\ 
	Conv (Id)            & 32x5x229@3x3    & 288             \\ 
	BatchNorm            & 32x5x229        & 128             \\ 
	Relu                 & 32x5x229        &                 \\ 
	Conv (Id)            & 32x3x227@3x3    & 9216            \\ 
	BatchNorm            & 32x3x227        & 128             \\ 
	Relu                 & 32x3x227        &                 \\ 
	MaxPool              & 32x3x113@1x2    &                 \\ 
	Dropout, p=0.25      & 32x3x113        &                 \\ 
	Conv (Id)            & 64x1x111@3x3    & 18432           \\ 
	BatchNorm            & 64x1x111        & 256             \\ 
	Relu                 & 64x1x111        &                 \\ 
	MaxPool              & 64x1x55@1x2     &                 \\ 
	Dropout, p=0.25      & 64x1x55         &                 \\ 
	Dense (Id)           & 512             & 1802240         \\ 
	BatchNorm            & 512             & 2048            \\ 
	Relu                 & 512             &                 \\ 
	Dropout, p=0.5       & 512             &                 \\ 
	Dense (Sigmoid)      & 88              & 45144           \\ 
\hline
	                     &                 & $\sum$ 1877880 
\end{tabular}%
}
\caption{The \ConvNet Architecture}
\label{table:convnet}
\end{table}
\begin{table}[htp]
\centering
\scalebox{0.6}{%
\begin{tabular}{llr}
	Layer                & Output          & No. of          \\ 
	Type                 & Dimensions      & Params          \\ 
\hline
	Input                & 1x5x229         &                 \\ 
	Conv (Id)            & 8x5x229@3x3     & 72              \\ 
	BatchNorm            & 8x5x229         & 32              \\ 
	Relu                 & 8x5x229         &                 \\ 
	Conv (Id)            & 8x3x227@3x3     & 576             \\ 
	BatchNorm            & 8x3x227         & 32              \\ 
	Relu                 & 8x3x227         &                 \\ 
	MaxPool              & 8x3x113@1x2     &                 \\ 
	Dropout, p=0.25      & 8x3x113         &                 \\ 
	Conv (Id)            & 8x1x111@3x3     & 576             \\ 
	BatchNorm            & 8x1x111         & 32              \\ 
	Relu                 & 8x1x111         &                 \\ 
	MaxPool              & 8x1x55@1x2      &                 \\ 
	Dropout, p=0.25      & 8x1x55          &                 \\ 
	Conv (Id)            & 8x1x53@1x3      & 192             \\ 
	BatchNorm            & 8x1x53          & 32              \\ 
	Relu                 & 8x1x53          &                 \\ 
	MaxPool              & 8x1x26@1x2      &                 \\ 
	Dropout, p=0.25      & 8x1x26          &                 \\ 
	Dense (Id)           & 16              & 3328            \\ 
	BatchNorm            & 16              & 64              \\ 
	Relu                 & 16              &                 \\ 
	Dropout, p=0.5       & 16              &                 \\ 
	Dense (Sigmoid)      & 23              & 391             \\ 
\hline
	                     &                 & $\sum$ 5327    
\end{tabular}%
}
\caption{The \SmallConvNet Architecture}
\label{table:small_convnet}
\end{table}
\begin{table}[htp]
\centering
\scalebox{0.6}{%
\begin{tabular}{llr}
	Layer                & Output          & No. of          \\ 
	Type                 & Dimensions      & Params          \\ 
\hline
	Input                & 1x256x256       &                 \\ 
	Conv (Id)            & 32x256x256@3x3  & 288             \\ 
	BatchNorm            & 32x256x256      & 128             \\ 
	Elu                  & 32x256x256      &                 \\ 
	Conv (Id)            & 32x256x256@3x3  & 9216            \\ 
	BatchNorm            & 32x256x256      & 128             \\ 
	Elu                  & 32x256x256      &                 \\ 
	MaxPool              & 32x128x128@2x2  &                 \\ 
	Conv (Id)            & 32x128x128@3x3  & 9216            \\ 
	BatchNorm            & 32x128x128      & 128             \\ 
	Elu                  & 32x128x128      &                 \\ 
	Conv (Id)            & 32x128x128@3x3  & 9216            \\ 
	BatchNorm            & 32x128x128      & 128             \\ 
	Elu                  & 32x128x128      &                 \\ 
	MaxPool              & 32x64x64@2x2    &                 \\ 
	Conv (Id)            & 64x64x64@3x3    & 18432           \\ 
	BatchNorm            & 64x64x64        & 256             \\ 
	Elu                  & 64x64x64        &                 \\ 
	Conv (Id)            & 64x64x64@3x3    & 36864           \\ 
	BatchNorm            & 64x64x64        & 256             \\ 
	Elu                  & 64x64x64        &                 \\ 
	MaxPool              & 64x32x32@2x2    &                 \\ 
	Conv (Id)            & 64x32x32@3x3    & 36864           \\ 
	BatchNorm            & 64x32x32        & 256             \\ 
	Elu                  & 64x32x32        &                 \\ 
	Conv (Id)            & 64x32x32@3x3    & 36864           \\ 
	BatchNorm            & 64x32x32        & 256             \\ 
	Elu                  & 64x32x32        &                 \\ 
	MaxPool              & 64x16x16@2x2    &                 \\ 
	Conv (Id)            & 128x16x16@3x3   & 73728           \\ 
	BatchNorm            & 128x16x16       & 512             \\ 
	Elu                  & 128x16x16       &                 \\ 
	Conv (Id)            & 128x16x16@3x3   & 147456          \\ 
	BatchNorm            & 128x16x16       & 512             \\ 
	Elu                  & 128x16x16       &                 \\
\end{tabular}%
\begin{tabular}{llr}
 
	Upscale              & 128x32x32       &                 \\ 
	Concat               & 192x32x32       &                 \\ 
	Conv (Id)            & 128x32x32@3x3   & 221184          \\ 
	BatchNorm            & 128x32x32       & 512             \\ 
	Elu                  & 128x32x32       &                 \\ 
	Conv (Id)            & 128x32x32@3x3   & 147456          \\ 
	BatchNorm            & 128x32x32       & 512             \\ 
	Elu                  & 128x32x32       &                 \\ 
	Upscale              & 128x64x64       &                 \\ 
	Concat               & 192x64x64       &                 \\ 
	Conv (Id)            & 64x64x64@3x3    & 110592          \\ 
	BatchNorm            & 64x64x64        & 256             \\ 
	Elu                  & 64x64x64        &                 \\ 
	Conv (Id)            & 64x64x64@3x3    & 36864           \\ 
	BatchNorm            & 64x64x64        & 256             \\ 
	Elu                  & 64x64x64        &                 \\ 
	Upscale              & 64x128x128      &                 \\ 
	Concat               & 96x128x128      &                 \\ 
	Conv (Id)            & 32x128x128@3x3  & 27648           \\ 
	BatchNorm            & 32x128x128      & 128             \\ 
	Elu                  & 32x128x128      &                 \\ 
	Conv (Id)            & 32x128x128@3x3  & 9216            \\ 
	BatchNorm            & 32x128x128      & 128             \\ 
	Elu                  & 32x128x128      &                 \\ 
	Upscale              & 32x256x256      &                 \\ 
	Concat               & 64x256x256      &                 \\ 
	Conv (Id)            & 32x256x256@3x3  & 18432           \\ 
	BatchNorm            & 32x256x256      & 128             \\ 
	Elu                  & 32x256x256      &                 \\ 
	Conv (Id)            & 32x256x128@3x3  & 9216            \\ 
	BatchNorm            & 32x256x128      & 128             \\ 
	Elu                  & 32x256x128      &                 \\ 
	Conv (Sigmoid)       & 1x256x88@1x41   & 1313            \\ 
\hline
	                     &                 & $\sum$ 964673  
\end{tabular}%
}
\caption{The \AUNet Architecture}
\label{table:aunet}
\end{table}

\end{document}